\documentclass[aps,prb,amsmath,amssymb,showpacs,reprint,
superscriptaddress
]{revtex4-1}

\usepackage{hyperref}
\usepackage{graphicx}
\usepackage{bm}

\begin{document}


\title{Manipulating the magnetic anisotropy of cobalt doped titanium dioxide by carrier accumulation}


\author{Bin Shao}
\affiliation{College of Information Technical Science, Nankai University, Tianjin 300071, China.}

\author{Min Feng}
\affiliation{School of Physics, Nankai University, Tianjin 300071, China.}

\author{Hong Liu}
\affiliation{Office of International Academic Exchanges, Nankai University, Tianjin 300071, China.}

\author{Jian Wu}
\affiliation{Physics Department, Tsinghua University, Beijing 100084, China.}


\author{Xu Zuo}
\email[Electronical Mail: ]{xzuo@nankai.edu.cn}
\affiliation{College of Information Technical Science, Nankai University, Tianjin 300071, China.}


\date{\today}

\begin{abstract}

Based on first-principles calculations, we predict that the magnetic anisotropy energy (MAE) of Co-doped TiO$_2$ sensitively depends on carrier accumulation. This magnetoelectric phenomenon provides a promising route to directly manipulate the magnetization direction of diluted magnetic semiconductor by external electric-fields. We calculate the band structures and reveal the origin of carrier-dependent MAE in \textit{k}-space. In fact, the carrier accumulation shifts the Fermi energy and regulates the competing contributions to MAE. The first-principles calculations provide a straightforward way to design spintronics materials with electrically controllable spin direction.

\end{abstract}


\maketitle


Diluted magnetic semiconductor (DMS), exhibiting both ferromagnetism and semiconducting properties, has been considered to be a promising candidate of spintronics. It provides a potential route to add the spin degree of freedom to conventional charge-based electronic devices, for example, adding magnetic recording capability to current semiconductor information processing unit. However, there are two major challenges, one is how to raise the Curie temperature ($T_{\text{C}}$), and the other is how to control the magnetization direction. Since the electric-field-induced room-temperature ferromagnetism has been demonstrated in cobalt-doped titanium dioxide (Co:TiO$_2$),\cite{Yamada27052011} where the ferromagnetic exchange is considered to be mediated by carriers and tuned by electric-field,  the left issue is how to manipulate the magnetization direction.

Conventionally, the magnetization direction is controlled by magnetic-field, which is unsuitable for ultrahigh-density magnetic storage and integration with electronic devices. Therefore, explorations aiming at manipulation of magnetization direction directly by electric-field have emerged. Early experiment \cite{Chiba15082003} reveals that the coercive force $H_{\text{C}}$ depends on the variation of carrier density managed by electric-field in Mn-doped InAs. This phenomenon hints that there is a correlation between the magnetic anisotropy and carrier density in DMS. Further work \cite{chiba2008magnetization} exhibits the rotation of magnetization direction by applying electric-field on Mn-doped GaAs, showing the direct connection between magnetic anisotropy and carrier density.

When an external electric-field is applied, there will be an accumulation of electrons or holes, yielding a shift of Fermi energy. Since magnetic anisotropy energy (MAE) is mainly determined by the band structure near the Fermi energy, the shift will impact the MAE. Early first-principles calculations of MAE in transition metal bulks and thin films have shown this effect by the so-called electron-filling technique, where electrons are added to or removed from the system under study. \cite{PhysRevB.41.11919, PhysRevB.47.14932, Wu1999498} We expect that carrier accumulation may also impact the MAE in DMS. Moreover, it will be easier to realize a carrier accumulation in DMS in experiment than magnetic metals.

In this letter, for Co:TiO$_2$, a typical DMS, our numerical calculations demonstrate that the carrier accumulation yields an oscillation of MAE. Instead of analyzing MAE within traditional single-ion anisotropy theory, a straightforward analysis based on detailed band structure near the Fermi energy is proposed to explain the dependence of MAE on carrier accumulation.

Based on density functional theory (DFT), first-principles calculations were carried out on Co-doped TiO$_2$ anatase using Perdew-Burke-Ernzerhof (PBE) parameterization \cite{PhysRevLett.77.3865} of generalized gradient approximation (GGA) as implemented in VASP package. \cite{PhysRevB.54.11169} The primitive anatase cell was fully optimized. Then, a $2\times2\times1$ supercell was created with one Ti atom substitued by Co (Co$_{0.0625}$Ti$_{0.9375}$O$_2$), and the atomic positions were allowed to relax. After the optimization, GGA+\textit{U} (GGA plus on-site Coulomb repulsion) approach \cite{PhysRevB.57.1505} was employed in electronic structure calculations. We applied extra Coulomb repulsion to Ti-\textit{d} orbital (2 eV) \cite{Ti_U} and Co-\textit{d} orbital (2 eV) \citep{PhysRevLett.100.256401}. The plane wave cut-off energy was 500 eV. The tetrahedron method with a $5\times5\times4$ \textit{k}-mesh grid was employed for the integration in Brillouin zone. The accuracy of electronic iterations was up to 10$^{-6}$ eV. The MAE, merely considering the contribution from spin-orbit coupling, was calculated as $\text{MAE}=E_{[100]} - E_{[001]}$, where $E_{[100]}$ and $E_{[001]}$ were the total energy with magnetization directions along [100] and [001], respectively. \cite{SOC} Carrier accumulation was simulated by modifying the total number of electrons, assuming a homogeneous background charge. The MAE is calculated for multiple charge configurations.

\begin{figure}
 \includegraphics{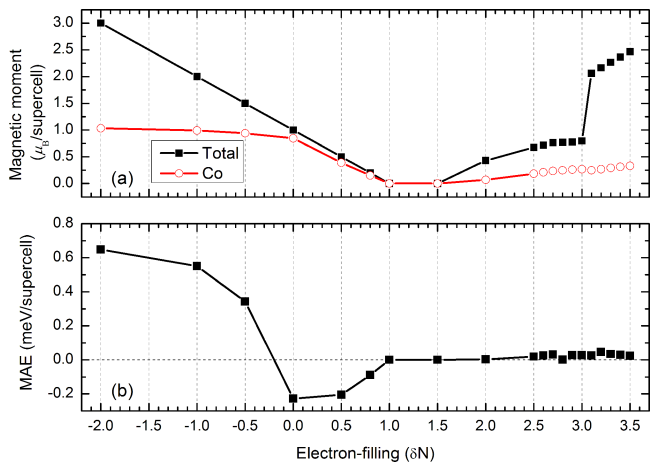}%
 \caption{\label{fig1}(Color online) (a) The total magnetic moment, magnetic moment on Co atom and (b) MAE of Co$_{0.0625}$Ti$_{0.9375}$O$_2$ depend on electron-filling.}
 \end{figure}

FIG.~\ref{fig1} shows the MAE and the magnetic moment as a function of the electron number added ($\delta N$), where negative $\delta N$ means that electrons are removed. It is obvious that both the MAE and the magnetic moment sensitively depend on electron-filling. In neutral Co:TiO$_2$, the MAE is negative, implying the anatase \textit{c}-plane is the easy plane of the magnetization. However, when electrons are removed, the easy-axis rotates out of the \textit{c}-plane, and the \textit{c}-axis becomes the easy-axis, and the magnitude of MAE increases steeply. In addition, when electrons are added, the magnitude of easy plane magnetic anisotropy begins to reduce and then becomes zero at $\delta N = 1.0$ and 1.5. With more electrons doped, the easy-axis aligns along the \textit{c}-axis, but the magnitude of MAE is weak, less than 0.1 meV. According to these results, the magnetization direction will switch from in-plane to perpendicular by varying carrier density, which can be achieved by carrier accumulation and tuned by external electric-field.

When $\delta N$ decreases from 0 to -2.0, the total magnetic moment of the supercell ($M_{\text{tot}}$) linearly increases from 1.0 to 3.0 $\mu_{\text{B}}$, however, the local magnetic moment on Co ($M_{\text{Co}}$) remains steady. In contrast, when $\delta N$ varies from 0 to 1.0, both $M_{\text{tot}}$ and $M_{\text{Co}}$ decrease to 0 $\mu_{\text{B}}$, indicating the system becomes non-magnetic. The nearly coincidence of the two curves suggests the magnetic moment mainly contributes from the Co atom at this range of $\delta N$. Further, we continue to increase $\delta N$ from 1.5 to 3.0, the system regains magnetic moments. Finally, when more than three electrons per supercell are added, the $M_{\text{tot}}$ drastically rises to 2.0 $\mu_{\text{B}}$ and then linearly increases to 2.5 $\mu_{\text{B}}$, while $M_{\text{Co}}$ levels off.

\begin{figure}
 \includegraphics{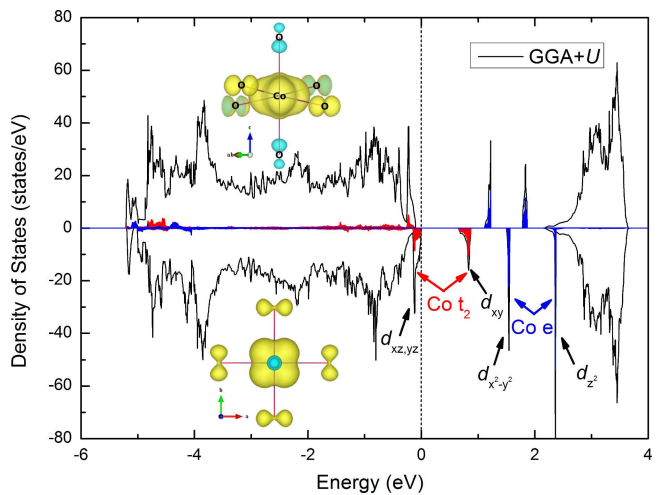}%
 \caption{\label{fig2}(Color online) Spin density maps (inset) and density of states for Co$_{0.0625}$Ti$_{0.9375}$O$_2$ with GGA+\textit{U} ($U_{\text{Ti}-d} = 2 $ eV, $U_{\text{Co}-d}= 2$ eV) approach, the vertical dash line at 0 eV is the Fermi energy. Solid line plot, total; red filled plot, Co-\textit{d} $t_2$; blue filled plot, Co-\textit{d} $e$, respectively.}
 \end{figure}

To understand the impact of electron-filling on magnetic moment and MAE, we first check the electronic structure of Co:TiO$_2$. The density of states (FIG.~\ref{fig2}) for neutral Co:TiO$_2$ with GGA+\textit{U} shows that the band gap of pristine TiO$_2$ is preserved, with the impurity bands lying in the gap. The well isolated impurity bands are mainly from the Co \textit{d}-orbitals, which hybrid with O \textit{p}-orbitals near the valence band maximum (VBM), and lead to slight magnetization of O atom as shown in the spin density map (FIG.~\ref{fig2}). The majority of Co $t_2$ manifold is completely occupied, while the minority of $t_2$ manifold is split into an occupied doublet ($d_{\text{xz,yz}}$) and an empty singlet ($d_{\text{xy}}$). The insulator ground state is different from the half-metallic ground state predicted by GGA.\cite{PhysRevB.65.161201,PhysRevB.67.144415,PhysRevB.73.035201} The total magnetic moment of the cell ($M_{\text{tot}} = 1.0$  $\mu_{\text{B}}$) and the spin density map (FIG.~\ref{fig2}) of Co atom with a shape of $d_{\text{xy}}$ orbital suggest a low-spin state, $(t^{\uparrow}_2)^{3}(t^{\downarrow}_2)^2$, which is consistent with previous beyond-DFT results. \cite{PhysRevB.65.161201,PhysRevB.73.035201}

\begin{figure}
 \includegraphics{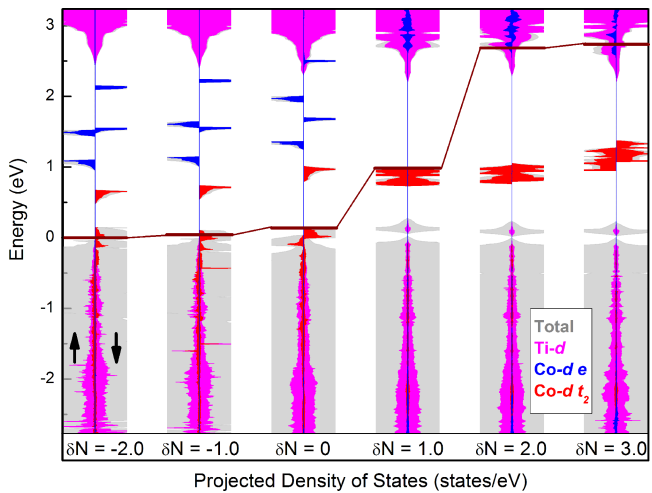}%
 \caption{\label{fig3}(Color online) The projected density of states (PDOS) of Co$_{0.0625}$Ti$_{0.9375}$O$_2$ with different $\delta{N}$. The PDOS is aligned by the deep O-\textit{s} ( $\sim-17 $ eV). Therefore, the electron-filling causes an obvious shift of the Fermi energy (horizontal wine line). Gray filled plot, total; magenta filled plot, Ti-\textit{d}; blue filled plot, Co-\textit{d} $e$; red filled plot, Co-\textit{d} $t_2$.}
 \end{figure}

The electron-filling causes an obvious shift of the Fermi energy (FIG.~\ref{fig3}), if the projected density of states with different $\delta N$ is aligned to the deep O \textit{s}-orbital. When one electron is added ($\delta N = 1.0$), the singlet ($d_{\text{xy}}$) in minority spin is occupied, and the system becomes non-magnetic. In addition, the Co-\textit{d} $e$ manifold is pushed into the conduction band (CB) of TiO$_2$ host. When two electrons are added, the Fermi energy shifts further into the host CB, and results in the so-called \textquotedblleft band-filling effect\textquotedblright, i.e., the host conduction band minimum (CBM) will be first occupied and then the $e$ manifold of Co will be partially filled. \cite{PhysRevB.72.035211} Similarly, when electrons are removed, they will be partially removed from the Co $t_2$. Therefore, $M_{\text{Co}}$ almost maintains a constant value with $\delta N < 0$ and $\delta N > 1.0$ [FIG.~\ref{fig2}(a)], which is called the negative-feedback charge regulation in Ref.~\onlinecite{raebiger2008charge}.

It is well-known that the spin-orbit coupling (SOC) interaction of 3\textit{d} transition metal elements is much weaker than the crystal-field split, and that MAE can be estimated by single-ion anisotropy theory.\cite{PTP.33.559,Yosida} However, the system becomes metallic when $\delta N < 0$ and $\delta N > 1.0$, and consequently, the charge state of Co is no longer well-defined under electron-filling. Then, the traditional single-ion anisotropy theory, where a well-defined charge state of magnetic ion is required, might not be suitable under this circumstance.

On the other hand, the MAE can be obtained by integrating the net contributions of the SOC interaction between the 3\textit{d} subbands in \textit{k}-space.\cite{PhysRev.58.909,PhysRevB.39.865,PhysRevB.47.14932} The contributions from degenerated and non-degenerated perturbations results in the first-order and second-order contributions, respectively. For the non-degenerated part, the contribution to MAE depends on the interaction between the occupied and empty states.\cite{PhysRevB.47.14932} It should be noted that the degenerated contribution could be as important as the non-degenerated part, although the degeneracy occurs only in a small portion of the Brillouin zone.\cite{PhysRevLett.87.216405} Thus, we compare the unperturbed band structures near the Fermi energy for different $\delta N$.

 \begin{figure}
 \includegraphics{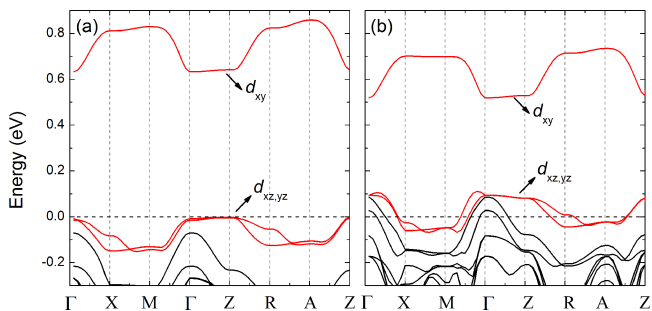}%
 \caption{\label{fig4}(Color online) Band structure of minority spin near the Fermi energy (horizontal dash line at 0 eV) along the edge of the irreducible Brillion zone (IBZ) without SOC, (a)$\delta N$ = 0 and (b) $\delta N$ = -1.0. The Co-\textit{d} bands are plotted in red.}
 \end{figure}

Due to the fully occupied $t_2$ manifold in majority spin and the low-spin configuration, the contribution from the spin-flip terms has been omitted.\cite{PhysRevB.47.14932,PhysRevB.39.865,0953-8984-10-14-012} We only plot band structures of minority spin near the Fermi energy for $\delta N = 0$ and -1.0 in FIG.~\ref{fig4}(a) and (b) without SOC, respectively. It is obvious that the $t_2$ manifold splits into a singlet ($d_{\text{xy}}$) and a doublet ($d_{\text{xz,yz}}$), due to the local $\text{D}_{\text{2d}}$ symmetry of the Co dopant. The doublet further splits along some directions, e.g. Z to R, due to the dispersion, when the translation symmetry is considered. As a result, the contribution to MAE can be divided into three categories: (i) the SOC interaction between occupied $d_{\text{xz,yz}}$ and empty $d_{\text{xy}}$, (ii) the SOC interaction between occupied $d_{\text{xz}}$($d_{\text{yz}}$) and empty $d_{\text{yz}}$($d_{\text{xz}}$), and (iii) the SOC interaction of the degenerated states inside the doublet $d_{\text{xz,yz}}$. The sign of MAE can be estimated by summing above three contributions.

For (i) the SOC interaction between occupied $d_{\text{xz,yz}}$ and empty $d_{\text{xy}}$ that have different magnetic quantum numbers, the perturbation is through $\bm{L_x}$ operator and yields a negative contribution.\cite{Wu1999498} For (ii) the SOC interaction between occupied $d_{\text{xz}}$($d_{\text{yz}}$) and empty $d_{\text{yz}}$($d_{\text{xz}}$) that share the same magnetic quantum numbers, the perturbation is through $\bm{L_z}$ operator and yields a positive contribution.\cite{Wu1999498} Note that the energy split between $d_{\text{xz}}$ and $d_{\text{yz}}$ is almost tenth of that between $d_{\text{xz,yz}}$ and $d_{\text{xy}}$. The contribution from (ii) can be 10 times larger than (i). For (iii) degenerated state, $\bm{L}$ is unquenched. In the subspace spanned by $\vert xz\rangle$ and $\vert yz\rangle$, the SOC Hamiltonian can be written as

\begin{equation}
\mathcal{H}=\lambda \bm{L}\cdot \bm{S}=\frac{\lambda }{2}\cos \theta 
\begin{pmatrix}
  0 & -i \\
  i & 0 
\end{pmatrix},
\end{equation}
where $\theta$ is the angle between $\bm{L}$ and $\bm{S}$. Our calculation shows that $\theta$ is also equal to the angle between spin direction and the \textit{c}-axis, since $\bm{L}$ is always along the \textit{c}-axis in the calculation. The eigenvalues of the Hamiltonian are

\begin{equation}
{{E}_{1}}=\frac{\left| \lambda  \right|}{2}\cos \theta ;{{E}_{0}}=-\frac{\left| \lambda  \right|}{2}\cos \theta.
\end{equation}

If there is only one electron in these states, e.g. $\delta N = -1.0$, the low-lying level $E_0$ will be occupied. As a result, the energy of the system depends on $\theta$ with the minimum at $\theta = 0$, i.e. the easy-axis is parallel to the \textit{c}-axis. Therefore, we conclude that, for degenerated sates, the contribution to MAE is positive.

With above analysis, the results of MAE for $\delta N = 0$ and $\delta N < 0$ can be explained qualitatively. For $\delta N = 0$, there is only perturbation of category (i), resulting in the moderate negative MAE. For $\delta N = -1.0$, there are all three contributions, where both categories (ii) and (iii) are positive contributions, and the magnitude of category (ii) is larger than (i). Thus, the MAE is positive. When more electrons are removed, the proportion of the positive contribution from (ii) increases, and the magnitude of MAE is larger than that of $\delta N = -1.0$.

\begin{figure}
 \includegraphics{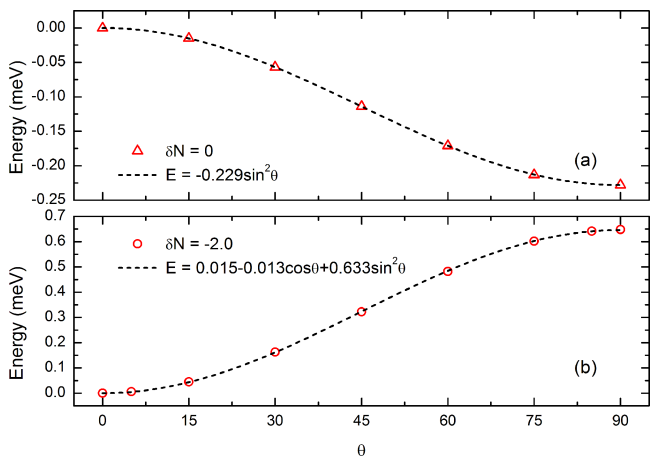}%
 \caption{\label{fig5}(Color online) The dependence of the total energy on the angle ($\theta$) between the quantum axis of spin and the anatase \textit{c}-axis, (a)$\delta N$ = 0 and (b) $\delta N$ = -2.0.}
 \end{figure}

The dependence of the total energy on the angle ($\theta$) between the quantum axis of spin and the anatase \textit{c}-axis (FIG.~\ref{fig5}) supports our argument quantitatively. The coefficient of $\cos \theta$, the first-order perturbation, indicates the contribution from degenerated perturbation. For $\delta N = 0$, it is less than $10^{-6}$ eV, implying the contribution from degenerated perturbation can be ignored. The sign of the coefficient of $\sin ^2 \theta$, the second order perturbation, includes the competing contributions from the non-degenerated perturbation.

For $\delta N > 1.0$, the MAE alters its sign to positive (the easy-axis parallels to the \textit{c}-axis) and decreases its magnitude to one order smaller than others. As shown in FIG.~\ref{fig3}, when $\delta N > 1.0$, the $e$ manifold becomes partially occupied. In fact, there will be a strong Jahn-Teller distortion for $d^7$ electronic configuration in low-spin state under octahedral crystal field. Consequently, the MAE will be weak, as Jahn-Teller effect increases the split between occupied and empty states and lowers down the total energy in general. In fact, without constraining the atomic positions, the total energy is lowered by 1.028 eV.

In summary, our first-principles calculations predict that the MAE in Co:TiO$_2$ can be controlled by carrier accumulation. This magnetoelectric phenomenon in this DMS system allows an effective manipulation of the magnetization direction directly by external electric-field or voltage. To interpret the behavior of MAE, the electronic structures with electron-filling are examined. A self-regulated feedback effect of local magnetic moment on Co has been discovered. We propose an explanation for MAE for metallic state where the charge state is not well-defined. The MAE is discussed in \textit{k}-space based on the band structure near the Fermi energy. Based on the perturbation method, the contribution to MAE has been divided into three categories. The shift of the Fermi energy caused by electron-filling regulates the contributions from different categories, and consequently determines the sign and magnitude of MAE.

\begin{acknowledgments}
 This research was sponsored by National Natural Science Foundation of China (Grant No. 10970499), National Basic Research Program of China (973 Program, Grant No. 2011CB606405).
\end{acknowledgments}


\begin{thebibliography}{23}
\bibitem [{Ti_()}]{Ti_U}%
  \BibitemOpen
  \href@noop {} {}\bibinfo {note} {The U parameter on Ti-d has been carefully
  checked from 0 to 6 eV. The conclusions in this wark are not sensitive to the
  choice of ${\mathrm{U}}_{\mathrm{Ti-d}}$.}\BibitemShut {Stop}%
\bibitem [{SOC()}]{SOC}%
  \BibitemOpen
  \href@noop {} {}\bibinfo {note} {The spin-orbit coupling was performed in
  non-self-consistent calculations based on the wavefunctions and charge
  distribution from self-consistent calculations without spin-orbit
  coupling.}\BibitemShut {Stop}%
\end{thebibliography}
\end{document}